\documentclass[proceedings]{JHEP3}
\usepackage{amsfonts}
\usepackage{amsmath}
\usepackage{epsfig}

\newbox\mybox

\newcommand\fverb{\setbox\mybox=\hbox\bgroup\verb}
\newcommand\fverbdo{\egroup\medskip\noindent\fbox{\unhbox\mybox}\ }
\newcommand\fverbit{\egroup\item[\fbox{\unhbox\mybox}]}
\conference{Hermitian versus non-Hermitian representations}
\abstract{We investigate four different types of representations of 
deformed canonical variables leading to generalized versions of Heisenberg's 
uncertainty relations resulting from noncommutative spacetime structures. We
demonstrate explicitly how the representations are related to each other and study
three characteristically different solvable models on these spaces, the harmonic oscillator, the manifestly
non-Hermitian Swanson model and an intrinsically noncommutative model 
with P\"{o}schl-Teller type potential. We provide an analytical expression for the metric
in terms of quantities specific to the generic solution procedure and show
that when it is appropriately implemented expectation values are 
independent of the particular representation.
A recently proposed  inequivalent representation resulting from Jordan twists is shown to lead to 
unphysical models. We suggest an anti-$\cal{PT}$-symmetric modification to overcome 
this shortcoming.}

\title{Hermitian versus non-Hermitian representations for minimal length
uncertainty relations}
\author{Sanjib Dey$^\bullet$, Andreas Fring$^\bullet$ and Boubakeur Khantoul$^\circ$ \\
$^\bullet$ Department of Mathematics, City University London,\\
$\,\,$ Northampton Square, London EC1V 0HB, UK\\
$^\circ$ Department of Physics, University of Jijel, BP 98, Ouled Aissa,
18000 Jijel, Algeria \\
E-mail: sanjib.dey.1@city.ac.uk, a.fring@city.ac.uk, bobphys@gmail.com}

\input{tcilatex}
\begin{document}

\section{Introduction}

Generalized versions of Heisenberg's uncertainty relations for deformed
canonical variables have attracted a considerable amount of attention \cite%
{Kempf1,Kempf2,AFBB,AFLGFGS,AFLGBB,DFG,Aschieri:2005zs,Gomes:2009tk} since
they lead to the interesting feature of minimal lengths and minimal momenta.
In three dimensions it was explicitly shown \cite{DFG} that the deformed
canonical variables are related to noncommutative spacetime structures and
the corresponding analogues of the creation and annihilation operators
satisfy $q$-deformed oscillator algebras \cite{Arik,Kulish:1990eh}. We will
focus here on a one dimensional version of a noncommutative space which
results as a decoupled direction from a three dimensional version as shown
in \cite{DFG}%
\begin{equation}
\lbrack X,P]=i\hbar \left( 1+\check{\tau}P^{2}\right) .  \label{one}
\end{equation}%
Here $\check{\tau}:=\tau /(m\omega \hbar )>0$ has the dimension of an
inverse squared momentum and $\tau ~$is therefore dimensionless. Our
intention is here to investigate different types of models for different
representations for the operators obeying these relations. We will compare
four representations, denoted as $\Pi _{(i)}$ with $i\in \{1,2,3,4\}$, for $%
X $ and $P$ in relation (\ref{one}) expressed in terms of the standard
canonical variables $x$ and $p$ satisfying $[x,p]=i\hbar $%
\begin{eqnarray}
X_{(1)} &=&(1+\check{\tau}p^{2})x,~P_{(1)}=p,\qquad \quad X_{(2)}=(1+\check{%
\tau}p^{2})^{1/2}x(1+\check{\tau}p^{2})^{1/2},~P_{(2)}=p,  \label{2} \\
X_{(3)} &=&x,~P_{(3)}=\frac{1}{\sqrt{\check{\tau}}}\tan \left( \sqrt{\check{%
\tau}}p\right) ,\quad X_{(4)}=ix(1+\check{\tau}p^{2})^{1/2},~P_{(4)}=-ip(1+%
\check{\tau}p^{2})^{-1/2}.~~~~~  \label{3}
\end{eqnarray}%
Representation $\Pi _{(1)}$ is most obvious and most commonly used, but
manifestly non-Hermitian with regard to the standard inner product. This is
adjusted in the Hermitian representation $\Pi _{(2)}$ obtained from $\Pi
_{(1)}$ by an obvious similarity transformation, i.e. $\Pi _{(2)}=(1+\check{%
\tau}p^{2})^{-1/2}\Pi _{(1)}(1+\check{\tau}p^{2})^{1/2}$.

Representation $\Pi _{(3)}$ is Hermitian in the standard sense, albeit less
evident. Apart from an additional term in $X_{(3)}$ commuting with $P_{(3)}$%
, it appeared already in \cite{Kempf2} where it was found to be a
representation acting on the quasiposition wave function. Below we
demonstrate that for some concrete models it is also related in a
non-obvious way to $\Pi _{(1)}$ by the transformations to be outlined in
section 2.

We have also a particular interest in representation $\Pi _{(4)}$ as it can
be constructed systematically from Jordan twists accompanied by an
additional rotation. In \cite{Castro:2011in} a closely related version of
this representation, which we denote by $\Pi _{(4^{\prime })}$, occurred
without the additional factors $i$ and $-i$ in $X_{(4)}$ and $P_{(4)}$,
respectively. However, it is easily checked that this is incorrect and does
not produce the commutation relations (\ref{one}), as instead this variant
produces a minus sign on the right hand side in front of the $\check{\tau}%
P^{2}$-term. One might consider that version of a noncommutative space,
which will, however, lead immediately to more severe problems such as a pole
in the metric etc. We will argue here further that the construction provided
in \cite{Castro:2011in} results in unphysical models and requires the
proposed adjustments.

We also note that representation $\Pi _{(4)}$ respects a different kind of $%
\mathcal{PT}$-symmetry. Whereas the $\mathcal{PT}$-symmetry $x\rightarrow -x$%
, $p\rightarrow p$, $i\rightarrow -i$ of the standard canonical variables is
inherited in a one-to-one fashion by the deformed variables in
representations $\Pi _{(i)}$ for $i=1,2,3$, i.e. $X_{(i)}\rightarrow
-X_{(i)} $, $P_{(i)}\rightarrow P_{(i)}$, $i\rightarrow -i$, it becomes an
anti-$\mathcal{PT}$-symmetry for $\Pi _{(4)}$, that is $X_{(i)}\rightarrow
X_{(i)}$, $P_{(i)}\rightarrow -P_{(i)}$, $i\rightarrow -i$. Both versions
are of course symmetries of the commutation relation (\ref{one}) and since
both of them are antilinear involutions, they may equally well be employed
to ensure the reality of spectra for operators respecting the symmetry \cite%
{EW,Bender:1998ke}.

We expect that in concrete models the physics, such as the expectation
values for observables, are independent of the representation. We will argue
here that this is indeed the case.

Our manuscript is organized as follows: In section 2 we provide a general
construction procedure which can be used to solve non-Hermitian models. In
section 3, 4 and 5 we discuss the harmonic oscillator, the Swanson model and
a model with P\"{o}schl-Teller type potential, respectively, in terms of the
aforementioned representations. Our conclusions are stated in section 6.

\section{A general construction procedure for solvable non-Hermitian
potentials}

Once a Hamiltonian for a potential system is formulated on a noncommutative
space it usually ceases to be of potential type. Our aim here is to find
exact solutions for the corresponding Schr\"{o}dinger equation. Let us first
explain the general method we are going to employ. It consists of four main
steps: In the first we convert the system to a potential one, in the second
we construct the explicit solution to that system as a function of the
energy eigenvalues $E$, in the third step we employ a quantization condition
by means of the choice of appropriate boundary conditions and in the final
step we have to construct an appropriate metric due to the fact that the
Hamiltonian might not be Hermitian.

We exploit the fact that for a large class of one dimensional models on
noncommutative spaces the Schr\"{o}dinger equation involving a Hamiltonian $%
H(p)$ in momentum space acquires the general form%
\begin{equation}
H(p)\psi (p)=E\psi (p)\qquad \Leftrightarrow \qquad -f(p)\psi ^{\prime
\prime }(p)+g(p)\psi ^{\prime }(p)+h(p)\psi (p)=E\psi (p),  \label{1}
\end{equation}%
with $f(p)$, $g(p)$, $h(p)$ being some model specific functions and $E$
denoting the energy eigenvalue. This version of the equation may be
converted to a potential system, see for instance \cite%
{Bhattacharjie:1962xr,Jana:2009xa}, 
\begin{equation}
\tilde{H}(q)\psi (q)=E\psi (q)\qquad \Leftrightarrow \qquad -\phi ^{\prime
\prime }(q)+V(q)\phi (q)=E\phi (q),  \label{pot}
\end{equation}%
when transforming simultaneously the wavefunction and the momentum, 
\begin{equation}
\psi (p)=e^{\chi (p)}\phi (p),\quad \chi (p)=\int \frac{f^{\prime }(p)+2g(p)%
}{4f(p)}dp,\qquad \text{and\qquad }q=\int f^{-1/2}(p)dp,  \label{psiq}
\end{equation}%
respectively. In terms of the original functions $f(p)$, $g(p)$ and $h(p)$,
as defined by equation (\ref{1}), the potential is of the form%
\begin{equation}
V(q)=\left. \frac{4g^{2}+3\left( f^{\prime }\right) ^{2}+8gf^{\prime }}{16f}-%
\frac{f^{\prime \prime }}{4}-\frac{g^{\prime }}{2}+h\right\vert _{q}.
\label{V}
\end{equation}%
At this stage one could simply compare with the literature on solvable
potentials in order to extract an explicit solution. However, as the
literature contains conflicting statements and ambiguous notations, we will
present here a simple and transparent construction method for the solutions
adopted from \cite{Bhattacharjie:1962xr,Levai:1989eaa}.Furthermore, the
quantities constructed in the next step occur explicitly in the expression
for the metric. For the purpose of constructing a solvable potential we
factorize the wavefunction $\phi (q)$ in (\ref{pot}) further into%
\begin{equation}
\phi (q)=v(q)F[w(q)]  \label{phi}
\end{equation}%
with as yet unknown functions $v(q)$, $w(q)$ and $F(w)$. This Ansatz
converts the potential equation back into a second order equation of the
type (\ref{1}), albeit for the function $F(w)$,%
\begin{equation}
F^{\prime \prime }(w)+Q(w)F^{\prime }(w)+R(w)F(w)=0,  \label{2nd}
\end{equation}%
where%
\begin{equation}
Q(w):=\frac{2v^{\prime }}{vw^{\prime }}+\frac{w^{\prime \prime }}{\left(
w^{\prime }\right) ^{2}}\qquad \text{and\qquad }R(w):=\frac{E-V(q)}{\left(
w^{\prime }\right) ^{2}}+\frac{v^{\prime \prime }}{v\left( w^{\prime
}\right) ^{2}}.  \label{QR}
\end{equation}%
Using the first relation in (\ref{QR}) we can express $v$ entirely in terms
of $w$ and $Q$%
\begin{equation}
v(q)=\left( w^{\prime }\right) ^{-1/2}\exp \left[ \frac{1}{2}\int^{w(q)}Q(%
\tilde{w})d\tilde{w}\right] .  \label{v}
\end{equation}%
With the help of this expression we eliminate $v$ from the second relation
in (\ref{QR}) and express the difference between the energy eigenvalue and
the potential as 
\begin{equation}
E-V(q)=\frac{w^{\prime \prime \prime }}{2w^{\prime }}-\frac{3}{4}\left( 
\frac{w^{\prime \prime }}{w^{\prime }}\right) ^{2}+\left( w^{\prime }\right)
^{2}R(w)-\frac{\left( w^{\prime }\right) ^{2}Q^{\prime }(w)}{2}-\frac{\left(
w^{\prime }\right) ^{2}Q^{2}(w)}{4}.  \label{EV}
\end{equation}%
Assuming now that $F$ as introduced in (\ref{phi}) is a particular special
function satisfying the second order differential equation (\ref{2nd}) with
known $Q(w)$ and $R(w)$, the only unknown quantity left on the right hand
side of (\ref{EV}) is $w(q)$. In the general pursuit of constructing
solvable potentials one then selects terms on the right hand side of (\ref%
{EV}) to match the constant $E$ which in turn fixes the function $w$. The
remaining terms on the right hand side must then compute to a meaningful
potential. For the case at hand it has to equal $V(q)$ as computed in (\ref%
{V}). Assembling everything one has therefore obtained an explicit form for $%
\phi (q)$ in (\ref{phi}) and hence $\psi (p)$, as given in (\ref{psiq}),
together with the energy eigenvalues $E$.

In the next step we need to implement the appropriate boundary conditions
and quantize $\psi (p)$ to a well-defined $L^{2}(\mathbb{R})$-function $\psi
_{n}(p)$ for discrete eigenvalues $E_{n}$.

What is left is to construct an appropriate metric, since some of our
Hamiltonians are non-Hermitian, either resulting from the fact that we use a
non-Hermitian representation or from the Hamiltonian being manifestly
non-Hermitian in the first place, or a combination of both. In any of those
cases we have to re-define the metric $\rho $ on our Hilbert space to $%
\left\langle \tilde{\psi}_{n}\right\vert \left. \psi _{n}\right\rangle
_{\rho }$ $:=\left\langle \tilde{\psi}_{n}\right\vert \left. \rho \psi
_{n}\right\rangle $. We could follow standard procedures as outlined in the
recent literature on non-Hermitian systems \cite%
{Urubu,Bender:1998ke,AliI,Benderrev,Alirev,BFGJ} as for instance to solve
the relation $\rho H\rho ^{-1}=H^{\dagger }$ for the operator $\rho $.
However, for the scenario outlined in this section we can present a closed
analytical formula. Assuming at this stage further that the function $F$, as
introduced in (\ref{phi}), is an orthonormal function we have%
\begin{equation}
\delta _{n,m}=\dint \varrho (w)F_{n}\left( w\right) F_{m}\left( w\right)
^{\ast }dw=\dint \varrho (p)e^{-2\func{Re}\chi (p)}\left\vert
v(p)\right\vert ^{-2}\frac{dw}{dp}\psi _{n}(p)\psi _{n}^{\ast }(p)dp,
\end{equation}%
such that the metric is read off as%
\begin{equation}
\rho (p)=\varrho (p)e^{-2\func{Re}\chi (p)}\left\vert v(p)\right\vert ^{-2}%
\frac{dw}{dp}.  \label{rho}
\end{equation}%
In the first integral we might need the additional metric $\varrho (w)$ in
case the special function\ $F\left( w\right) $ is not taken to be
orthonormal. All quantities on the right hand side are explicitly known at
this point of the construction allowing us to compute $\rho (p)$ directly.
We note that that the positivity of the metric in entirely governed by $%
dw/dp $.

\section{The harmonic oscillator in different representations}

At first we will consider the harmonic oscillator in different
representations 
\begin{equation}
H_{(i)}=\frac{P_{(i)}^{2}}{2m}+\frac{m\omega ^{2}}{2}X_{(i)}^{2},\qquad 
\text{for }i=1,2,3,4,  \label{HD1}
\end{equation}%
with a particular focus on $\Pi _{(4)}$, which has not been dealt with so
far. In principle the solutions for $\Pi _{(1)}$ are known, but it is
instructive to consider here briefly how they emerge in the above scheme. In
terms of the standard canonical variables the Hamiltonian reads%
\begin{equation}
H_{(1)}(p)=\frac{p^{2}}{2m}+\frac{m\omega ^{2}}{2}\left( x^{2}+\check{\tau}%
p^{2}x^{2}+\check{\tau}xp^{2}x+\check{\tau}^{2}p^{2}xp^{2}x\right) ,\qquad 
\text{for }p\in \mathbb{R}.  \label{H1}
\end{equation}%
With $x=i\hbar \partial _{p}$, the corresponding Schr\"{o}dinger equation in
momentum space acquires the general form (\ref{1}), where we identify%
\begin{equation}
f(p)=\frac{m\omega ^{2}\hbar ^{2}}{2}(1+\check{\tau}p^{2})^{2},\quad
g(p)=-\tau \hbar \omega p(1+\check{\tau}p^{2}),\quad \text{and\quad }h(p)=%
\frac{p^{2}}{2m}.
\end{equation}%
Then the equations (\ref{psiq}) and (\ref{V}) convert this into an equation
for a potential system with Hamiltonian $\tilde{H}_{(1)}(q)$ 
\begin{equation}
\psi (p)=\phi (p),\quad q=\sqrt{\frac{2}{\tau \omega \hbar }}\arctan \left( 
\sqrt{\check{\tau}}p\right) ,\quad \text{and}\quad V(q)=\frac{\hbar \omega }{%
2\tau }\tan ^{2}\left( \sqrt{\frac{\tau \omega \hbar }{2}}q\right) .
\end{equation}%
The $\tan ^{2}$-potential is well known to be solvable, which is explicitly
seen as follows. Assuming that $F(w)$ is an associated Legendre polynomial $%
P_{\nu }^{\mu }(w)$ we identify from the defining differential equation for
these functions, see e.g. \cite{Grad}, the coefficient functions in (\ref%
{2nd}) as%
\begin{equation}
Q(w)=\frac{2w}{w^{2}-1}\qquad \text{and\qquad }R(w)=\frac{\nu (\nu +1)}{%
1-w^{2}}-\frac{\mu ^{2}}{\left( 1-w^{2}\right) ^{2}}.  \label{QRx}
\end{equation}%
Then equation (\ref{EV}) acquires the form%
\begin{equation}
E-\frac{\hbar \omega }{2\tau }\tan ^{2}\left( \sqrt{\frac{\tau \omega \hbar 
}{2}}q\right) =\left( w^{\prime }\right) ^{2}\left( \frac{\nu ^{2}+\nu +1}{%
1-w^{2}}+\frac{w^{2}-\mu ^{2}}{\left( 1-w^{2}\right) ^{2}}\right) -\frac{%
3\left( w^{\prime \prime }\right) ^{2}}{4\left( w^{\prime }\right) ^{2}}+%
\frac{w^{\prime \prime \prime }}{2w^{\prime }}  \label{E}
\end{equation}%
for the unknown function $w(q)$ and constant $E$. Assuming that the first
term on the right hand side gives rise to a constant, i.e. $\left( w^{\prime
}\right) ^{2}/(1-w^{2})=c\in \mathbb{R}^{+}$, we obtain $w(q)=\sin (\sqrt{c}%
q)$ as solution of the latter equation. This function solves (\ref{E}) with
the identifications%
\begin{equation}
E=\frac{\tau \omega \hbar }{8}(1+2\nu )^{2}-\frac{\hbar \omega }{2\tau }%
,\quad \quad c=\frac{\tau \omega \hbar }{2},\quad \quad \text{and\quad \quad 
}\mu =\mu _{\pm }=\pm \frac{1}{\tau }\sqrt{1+\frac{\tau ^{2}}{4}}.
\label{mu}
\end{equation}%
It remains to compute $v(q)$, which results from (\ref{v}), such that all
quantities assembled yield $\phi (q)$ in (\ref{phi}) as%
\begin{equation}
\phi (q)=\sqrt{\cos (\sqrt{\tau \omega \hbar /2}q)}P_{\nu }^{\mu _{\pm }}%
\left[ \sin (\sqrt{\tau \omega \hbar /2}q)\right] .  \label{qphi}
\end{equation}%
Hence with (\ref{psiq}) we obtain finally a solution to the Schr\"{o}dinger
equation in momentum space involving the Hamiltonian $H_{(1)}(x,p)$%
\begin{equation}
\psi (p)=\frac{1}{(1+\check{\tau}p^{2})^{1/4}}P_{\nu }^{\mu _{\pm }}\left( 
\frac{\sqrt{\check{\tau}}p}{\sqrt{1+\check{\tau}p^{2}}}\right) .
\label{fast}
\end{equation}%
At this stage the constant $\nu $ is still unspecified. Implementing now the
final step, the boundary conditions $\lim_{p\rightarrow \pm \infty }\psi (p)$
$=0$ yields the quantization condition for the energy. Using the property $%
\lim_{z\rightarrow \pm 1}P_{n-m}^{m}\left( z\right) =$ $0$ for $n\in \mathbb{%
N}$, $m<0$ we need to chose $\mu _{-}$ in (\ref{fast}), such that $\nu
=n+1/\tau \sqrt{1+\tau ^{2}/4}$. Therefore the asymptotically vanishing
eigenfunctions become%
\begin{equation}
\psi _{n}(p)=\frac{1}{\sqrt{N_{n}}}\frac{1}{(1+\check{\tau}p^{2})^{1/4}}%
P_{n-\mu _{-}}^{\mu _{-}}\left( \frac{\sqrt{\check{\tau}}p}{\sqrt{1+\check{%
\tau}p^{2}}}\right) ,  \label{310}
\end{equation}%
with corresponding energy eigenvalues 
\begin{equation}
E_{n}=\omega \hbar \left( \frac{1}{2}+n\right) \sqrt{1+\frac{\tau ^{2}}{4}}+%
\frac{\tau \omega \hbar }{4}(1+2n+2n^{2}),  \label{En}
\end{equation}%
and normalization constant $N_{n}$. The expression for $E_{n}$ agrees
precisely with the one previously obtained in \cite{Kempf2,DFG} by different
means. The corresponding eigenfunctions $\psi _{n}(p)$ are clearly $L^{2}(%
\mathbb{R})$-function, but since $H_{(1)}$ is non-Hermitian we do not expect
them to be orthonormal. Noting that%
\begin{equation}
\delta _{n,m}=\frac{1}{\sqrt{N_{n}N_{m}^{\ast }}}\dint%
\nolimits_{-1}^{1}P_{n-\mu _{-}}^{\mu _{-}}\left( w\right) \left[ P_{m-\mu
_{-}}^{\mu _{-}}\left( w\right) \right] ^{\ast }dw,  \label{tPP}
\end{equation}%
with normalization constant $N_{n}:=\dint\nolimits_{-1}^{1}\left\vert
P_{n-\mu _{-}}^{\mu _{-}}\left( z\right) \right\vert ^{2}dz$, we use $w=%
\sqrt{\check{\tau}}p/\sqrt{1+\check{\tau}p^{2}}$ to compute the metric from (%
\ref{rho}). We obtain $\rho (p)=\sqrt{\check{\tau}}\left( 1+\check{\tau}%
p^{2}\right) ^{-1}$, which apart from an irrelevant overall factor $\sqrt{%
\check{\tau}}$ is the same as the operator obtained from solving the
relations $\rho H\rho ^{-1}=H^{\dagger }$ as previously reported in \cite%
{AFBB,DFG}.

Since by (\ref{2}) it follows immediately that $H_{(2)}=\rho
^{1/2}H_{(1)}\rho ^{-1/2}$, the solutions for the Hermitian Hamiltonian $%
H_{(2)}$ are easily obtained from those for $H_{(1)}$ as $\rho ^{-1/2}\psi
_{n}$ with identical energy eigenvalues (\ref{En}).

For the representation $\Pi _{(3)}$ we notice that the associated
Hamiltonian $H_{(3)}(p)$ is just a rescaled version of the Hamiltonian $%
\tilde{H}_{(1)}(q)$, i.e. $H_{(3)}(p)=$ $\tilde{H}_{(1)}(q=p\sqrt{2/m}/\hbar
\omega )$ with $-\pi /2\sqrt{\check{\tau}}\leq p\leq \pi /2\sqrt{\check{\tau}%
}$. Thus the solution for the corresponding Schr\"{o}dinger equation is
simply $\phi (q=p\sqrt{2/m}/\hbar \omega )$. The metric results to be simply
an overall constant factor $\rho (p)=\sqrt{\check{\tau}}$, which is
consistent with the fact that $\Pi _{(3)}$ is Hermitian with regard to the
standard inner product.

Leaving the aforementioned problems for $\Pi _{(4^{\prime })}$ aside, we may
still consider whether it might yield a physically meaningful Hamiltonian.
In terms of the standard canonical variables we obtain 
\begin{equation}
H_{(4^{\prime })}(p)=\frac{p^{2}}{2m(1+\check{\tau}p^{2})}+\frac{m\omega ^{2}%
}{2}\left( x^{2}+\check{\tau}x^{2}p^{2}-i\hbar \check{\tau}xp\right) .
\end{equation}%
In momentum space the corresponding Schr\"{o}dinger equation is of the form (%
\ref{1}) with%
\begin{equation}
f(p)=\frac{m\hbar ^{2}\omega ^{2}}{2}(1+\check{\tau}p^{2}),\quad g(p)=-\frac{%
3}{2}\tau \hbar \omega p,\quad \text{and\quad }h(p)=\frac{p^{2}}{2m}(1+%
\check{\tau}p^{2})^{-1}-\frac{\tau \hbar \omega }{2}.  \label{fgh}
\end{equation}%
Then equations (\ref{psiq}) and (\ref{V}) yield%
\begin{equation}
\psi (p)=(1+\check{\tau}p^{2})^{-1/2}\phi (p),~q=\sqrt{\frac{2}{\tau \omega
\hbar }}\func{arcsinh}\left( \sqrt{\check{\tau}}p\right) ,~V(q)=\frac{\hbar
\omega }{2\tau }\tanh ^{2}\left( \sqrt{\frac{\tau \omega \hbar }{2}}q\right)
.  \label{psiV}
\end{equation}%
With the same assumption on $F(w)$ as made previously we obtain again the
relation (\ref{E}) with the difference that the $\tan ^{2}$-potential on the
left hand side is replaced by a $\tanh ^{2}$-potential. We may produce the
latter potential by assuming $\left( w^{\prime }\right) ^{2}/(1-w^{2})=-c$,
for $c\in \mathbb{R}^{+}$, which is solved by $w(q)=i\sinh (\sqrt{c}q)$.
However, the resulting energy eigenvalues $E=\hbar \omega /2\tau -c/4(1+2\nu
)^{2}$ are not bounded from below, which renders the Hamiltonian $%
H_{(4^{\prime })}$ as unphysical.

Using instead $H_{(4)}$ yields the same version of the Schr\"{o}dinger
equation, but all functions in (\ref{fgh}) are all replaced with an overall
minus sign. The corresponding quantities in (\ref{psiV}) are to be replaced
by $\psi (p)=(1+\check{\tau}p^{2})^{-1/2}P_{m-\mu _{-}}^{\mu _{-}}\left( -i%
\sqrt{\check{\tau}}p\right) $ with $-i/\sqrt{\check{\tau}}\leq p\leq i/\sqrt{%
\check{\tau}}$, the parameter $q$ needs to be multiplied by $-i$ and in the
potential the $\tanh ^{2}$ becomes a $\tan ^{2}$. Then the energy spectrum
becomes physically meaningful, being identical to (\ref{En}). The metric
results to $\rho (p)=-i\sqrt{\check{\tau}}\left( 1+\check{\tau}p^{2}\right)
^{1/2}$ in this case.

With the explicit solutions we may now verify that the expectation values
are indeed the same for all representations. For an arbitrary function $%
F\left( P_{(i)},X_{(i)}\right) $ we compute a universal expression 
\begin{equation}
\left\langle \psi _{(i)}\right\vert F\left( P_{(i)},X_{(i)}\right) \left.
\psi _{(i)}\right\rangle _{\rho _{(i)}}=\frac{1}{N}\dint\nolimits_{-1}^{1}F%
\left[ \frac{z}{\sqrt{\check{\tau}(1-z^{2})}},i\hbar \sqrt{\check{\tau}%
(1-z^{2})}\partial _{z}\right] \left\vert P_{m-\mu _{-}}^{\mu _{-}}\left(
z\right) \right\vert ^{2}dz,  \label{exHO}
\end{equation}%
for $i=1,2,3,4$. In particular we have $\left\langle \psi _{(i)}\right\vert
H_{(i)}\left. \psi _{(i)}\right\rangle _{\rho _{(i)}}=E_{n}$, $\left\langle
\psi _{(i)}\right\vert P_{(i)}\left. \psi _{(i)}\right\rangle _{\rho
_{(i)}}=0$.

\section{The Swanson model in different representations}

Let us next consider a model which is a widely studied \cite{MGH} solvable
prototype example to investigate non-Hermitian systems, the so-called
Swanson model \cite{Swanson}. On a noncommutative space it reads%
\begin{eqnarray}
H_{(i)} &=&\hbar \omega \left( A_{(i)}^{\dagger }A_{(i)}+\frac{1}{2}\right)
+\alpha A_{(i)}A_{(i)}+\beta A_{(i)}^{\dagger }A_{(i)}^{\dagger }\quad ~~~~~%
\text{for }i=1,2,3,4, \\
&=&\frac{\hbar \omega (1-\tau )-\alpha -\beta }{2m\hbar \omega }P_{(i)}^{2}+%
\frac{\Omega m\omega }{2\hbar }X_{(i)}^{2}+i\left( \frac{\alpha -\beta }{%
2\hbar }\right) \left( X_{(i)}P_{(i)}+P_{(i)}X_{(i)}\right) ,
\end{eqnarray}%
with $A_{(j)}=\left( m\omega X_{(j)}+iP_{(j)}\right) /\sqrt{2m\hbar \omega }$%
, $A_{(j)}^{\dagger }=\left( m\omega X_{(j)}-iP_{(j)}\right) /\sqrt{2m\hbar
\omega }$ and $\Omega :=\alpha +\beta +\hbar \omega $, $\alpha ,\beta \in 
\mathbb{R}$ with dimension of energy. Evidently for the standard inner
product we have in general $H_{(i)}\neq H_{(i)}^{\dagger }$ when $\alpha
\neq \beta $; even for $\tau =0$. Let us now study this model for the
different types of representations. Starting with $\Pi _{(1)}$, we obtain
the Schr\"{o}dinger equation in momentum space once again in the form of (%
\ref{1}), with%
\begin{eqnarray}
f(p) &=&\frac{m\hbar \omega \Omega }{2}(1+\check{\tau}p^{2})^{2},\quad
g(p)=(\beta -\alpha -\tau \Omega )p(1+\check{\tau}p^{2}), \\
h(p) &=&\frac{\beta -\alpha }{2}-\frac{\tau (\alpha -\beta +\hbar \omega
)+\alpha +\beta -\hbar \omega }{2hm\omega }p^{2}.  \notag
\end{eqnarray}%
Then equations (\ref{psiq}) and (\ref{V}) yield%
\begin{eqnarray}
\psi (p) &=&(1+\check{\tau}p^{2})^{\frac{(\beta -\alpha )}{2\tau \Omega }%
}\phi (p),\quad q=\sqrt{\frac{2}{\tau \Omega }}\arctan \left( \sqrt{\check{%
\tau}}p\right) ,  \label{psiqq} \\
V_{S\tan }(q) &=&\frac{(1-\tau )\hbar ^{2}\omega ^{2}-\tau \hbar \omega
(\alpha +\beta )-4\alpha \beta }{2\tau \Omega }\tan ^{2}\left( \sqrt{\frac{%
\tau \Omega }{2}}q\right) .  \label{Vq}
\end{eqnarray}%
Notice that we obtain again a $\tan ^{2}$-potential, albeit with different
constants involved. Using therefore as in the previous subsection the
assumption that $F(w)$ is an associated Legendre polynomial, we compute with
(\ref{QRx}) the equation (\ref{E}) with the left hand side replaced by $%
E-V_{S\tan }(q)$. With the same assumption on the function $w$, namely $%
\left( w^{\prime }\right) ^{2}/(1-w^{2})=c\in \mathbb{R}^{+}$, we obtain $%
w(q)=\sin (\sqrt{c}q)$ albeit now with $q$ taken from (\ref{psiqq}). The
equivalent to equation (\ref{E}) then yields%
\begin{eqnarray}
E &=&\frac{\tau \Omega }{8}(1+2\nu )^{2}+\frac{4\alpha \beta +\tau \hbar
\omega (\alpha +\beta )+\hbar ^{2}(\tau -1)\omega ^{2}}{2\tau \Omega },\quad
c=\frac{\tau \Omega }{2},  \label{EE} \\
\mu _{\pm } &=&\pm \frac{\sqrt{4\left( \hbar ^{2}\omega ^{2}-4\alpha \beta
\right) +\tau \Omega (\tau \Omega -4\hbar \omega )}}{2\tau \Omega }.
\label{mumu}
\end{eqnarray}%
Since $\phi (p)$ takes on the same form as in (\ref{psiq}) we obtain 
\begin{equation}
\psi _{n}(p)=\frac{1}{\sqrt{N_{n}}}(1+\check{\tau}p^{2})^{\frac{\beta
-\alpha }{2\tau \Omega }-\frac{1}{4}}P_{n-\mu _{-}}^{\mu _{-}}\left( \frac{%
\sqrt{\check{\tau}}p}{\sqrt{1+\check{\tau}p^{2}}}\right) ,  \label{psi}
\end{equation}%
as a solution to the Schr\"{o}dinger equation in momentum space involving
the Hamiltonian $H_{(1)}(p)$ for $p\in \mathbb{R}$. We have used the same
condition for the asymptotics of the wavefunction as stated before (\ref{310}%
), such that the energy eigenvalues become%
\begin{equation}
E_{n}=\frac{1}{4}\left[ (\tau +2n\tau +2n^{2}\tau )\Omega +(2n+1)\sqrt{%
4\left( \hbar ^{2}\omega ^{2}-4\alpha \beta \right) +\tau \Omega (\tau
\Omega -4\hbar \omega )}\right] .  \label{Enn}
\end{equation}%
Notice that in the commutative limit $\tau \rightarrow 0$ we recover the
well-known \cite{Swanson,MGH} expression for the energy $E_{n}=(n+1/2)\sqrt{%
\hbar ^{2}\omega ^{2}-4\alpha \beta }$. However, we find a discrepancy with
the results reported in \cite{Jana:2009xa} when taking the parameter $\gamma 
$ in there to zero. The authors do not state any quantization condition, but
besides that we can also not verify that the reported expression indeed
satisfies the relevant Schr\"{o}dinger equation.

In figure 1 we depict the onsets of the exceptional points as a function of
the parameters $\alpha $ and $\beta $ with the remaining parameters fixed.
We notice that for small values of $\alpha $ the domain for which the energy
is real is usually reduced, i.e. a model which still has real energy
eigenvalues on the standard space might develop complex eigenvalues on the
noncommutative space of the type (\ref{1}), e.g. for $\alpha =2$, $\beta
=0.1 $ we read off from the figure that $E_{n}(\tau =0)\in \mathbb{R}$
whereas $E_{n}(\tau =0.5)\notin \mathbb{R}$. In contrast, for larger values
of $\alpha $ complex eigenvalues might become real again once the model is
put onto the space of the type (\ref{1}), e.g. for $\alpha =15$, $\beta =0.1$
we find $E_{n}(\tau =0)\notin \mathbb{R}$ and $E_{n}(\tau =0.5)\in \mathbb{R}
$. Notice that the condition $\left( \hbar ^{2}\omega ^{2}-4\alpha \beta
\right) >\tau \Omega (\tau \Omega /4-\hbar \omega )$ which governs the
reality of the energy in (\ref{Enn}) is the same which controls the ${%
\mathcal{PT}}$-symmetry of the wavefunction $\psi (p)$, which is broken once 
$\mu _{-}\notin \mathbb{R}$.

\begin{figure}[h]
\centering   \includegraphics[width=12cm]{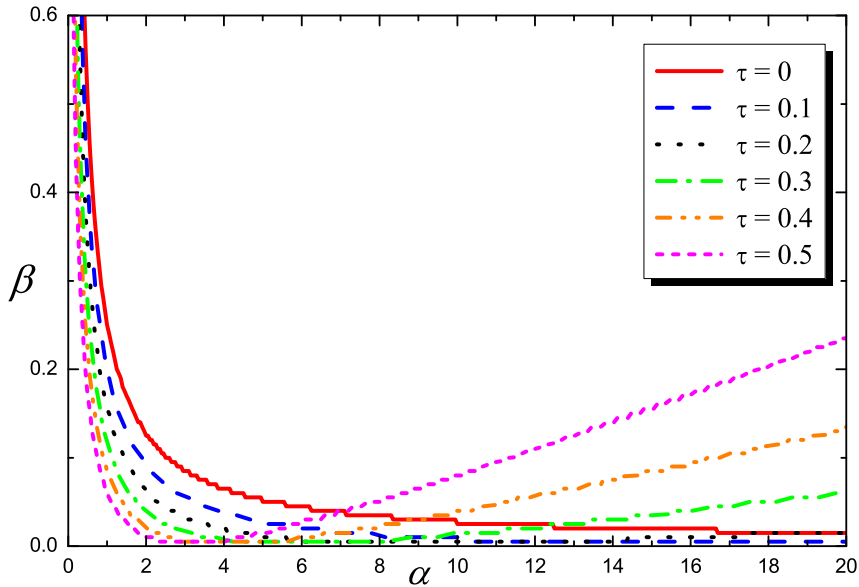}
\caption{Domain of spontaneously broken (above the curve) and unbroken
(below the curve) ${\mathcal{PT}}$-symmetry for the Swanson model with $%
\protect\omega =1$, $\hbar =1$, $m=1$ and different values of $\protect\tau $%
.}
\label{F1}
\end{figure}
With the help of (\ref{rho}) the metric is now computed to $\rho (p)=\sqrt{%
\check{\tau}}\left( 1+\check{\tau}p^{2}\right) ^{\frac{\alpha -\beta -\tau
\Omega }{\tau \Omega }}$.

Once again the solution for $H_{(2)}$ is $\rho ^{-1/2}\psi _{n}$ with energy
eigenvalues (\ref{Enn}) due to $H_{(2)}=\rho ^{1/2}H_{(1)}\rho ^{-1/2}$.

Next we consider the representation $\Pi _{(3)}$. The Schr\"{o}dinger
equation in momentum space acquires the general form of (\ref{1}), with%
\begin{eqnarray}
f(p) &=&\frac{m\hbar \omega \Omega }{2},\quad g(p)=\frac{\beta -\alpha }{%
\sqrt{\check{\tau}}}\tan \left( \sqrt{\check{\tau}}p\right) , \\
h(p) &=&\frac{\hbar \omega }{2}+\frac{\beta -\alpha -\hbar \omega }{2}\sec
^{2}\left( \sqrt{\check{\tau}}p\right) +\frac{\hbar \omega -\alpha -\beta }{%
2\tau }\tan ^{2}\left( \sqrt{\check{\tau}}p\right) .  \notag
\end{eqnarray}%
In this case the equations (\ref{psiq}) and (\ref{V}) yield%
\begin{equation}
\psi (p)=\left[ \cos \left( \sqrt{\check{\tau}}p\right) \right] ^{\frac{%
\alpha -\beta }{\tau \Omega }}\phi (p),\quad q=\sqrt{\frac{2}{m\hbar \omega
\Omega }}p,\quad V(q)=V_{S\tan }(q).
\end{equation}%
Notice that in the $q$-variables the potential obtained is exactly the same
as the one previously computed (\ref{Vq}) for representation $\Pi _{(3)}$.
Thus we obtain the same equation (\ref{EE}) and (\ref{mumu}) for the energy
and the parameter $\mu $, respectively. However, the corresponding
wavefunctions differ, resulting in this case to 
\begin{equation}
\psi _{n}(p)=\frac{1}{\sqrt{N_{n}}}\left[ \cos \left( \sqrt{\check{\tau}}%
p\right) \right] ^{\frac{\alpha -\beta }{\tau \Omega }+\frac{1}{2}}P_{n-\mu
_{-}}^{\mu _{-}}\left[ \sin \left( \sqrt{\check{\tau}}p\right) \right] ,
\end{equation}%
for $-\pi /2\sqrt{\check{\tau}}\leq p\leq \pi /2\sqrt{\check{\tau}}$. We
compute $\rho (p)=\sqrt{\check{\tau}}\left[ \cos \left( \sqrt{\check{\tau}}%
p\right) \right] ^{\frac{2(\beta -\alpha )}{\tau \Omega }}$ from (\ref{rho})
as relevant metric. Notice that $\rho (p)$ reduces to the standard metric
for $\alpha =\beta $ reflecting the fact that $H_{(3)}$ is Hermitian for
these values.

Since representation $\Pi _{(4^{\prime })}$ was identified as being
unphysical in the previous subsection, it is clear that this will also be
the case for the Swanson model and we will therefore not treat it any
further here.

For $\Pi _{(4)}$ the Schr\"{o}dinger equation in momentum space is also of
the form (\ref{1}), with%
\begin{eqnarray}
f(p) &=&\frac{m\hbar \omega \Omega }{2}(1+\check{\tau}p^{2}),\quad
g(p)=p\left( \beta -\alpha +\frac{3}{2}\tau \Omega \right) , \\
h(p) &=&\frac{1}{2(1+\check{\tau}p^{2})}\left\{ (\beta -\alpha +\tau \Omega
)+\frac{p^{2}}{m\hbar \omega }\left[ \alpha +\beta -\hbar \omega +\tau
(2\beta -2\alpha +\hbar \omega )+\tau ^{2}\Omega \right] \right\} .~~~~ 
\notag
\end{eqnarray}%
In this case the equations (\ref{psiq}) and (\ref{V}) yield%
\begin{equation}
\psi (p)=(1+\check{\tau}p^{2})^{\frac{\alpha -\beta }{2\tau \Omega }-\frac{1%
}{2}}\phi (p),\quad q=-i\sqrt{\frac{2}{\tau \Omega }}\func{arcsinh}\left( 
\sqrt{\check{\tau}}p\right) ,\quad V(q)=V_{S\tan }(q).
\end{equation}%
Notice that in the $q$-variables the potential obtained is exactly the same
as the one previously computed (\ref{Vq}) for representation $\Pi _{(3)}$.
Thus we obtain the same equations (\ref{EE}) and (\ref{mumu}) for the energy
eigenvalues and the parameter $\mu $, respectively. However, the final
wavefunction differs, resulting, after imposing the boundary conditions, to 
\begin{equation}
\psi _{n}(p)=\frac{1}{\sqrt{N_{n}}}(1+\check{\tau}p^{2})^{\frac{\alpha
-\beta }{2\tau \Omega }-\frac{1}{4}}P_{n-\mu _{-}}^{\mu _{-}}\left( -i\sqrt{%
\check{\tau}}p\right) ,
\end{equation}%
with $-i/\sqrt{\check{\tau}}\leq p\leq i/\sqrt{\check{\tau}}$. Now we
evaluate $\rho (p)=-i\sqrt{\check{\tau}}\left( 1+\check{\tau}p^{2}\right) ^{%
\frac{\beta -\alpha }{\tau \Omega }+\frac{1}{2}}$as metric from our general
formula (\ref{rho}).

We compute again the expectation values for some arbitrary function $F\left(
P_{(i)},X_{(i)}\right) $ in all four representations 
\begin{equation}
\left\langle \psi _{(i)}\right\vert F\left( P_{(i)},X_{(i)}\right) \left.
\psi _{(i)}\right\rangle _{\rho _{(i)}}=\frac{1}{N}\dint\nolimits_{-1}^{1}F%
\left[ \frac{z}{\sqrt{\check{\tau}(1-z^{2})}},i\hbar \sqrt{\check{\tau}%
(1-z^{2})}\partial _{z}\right] \left\vert P_{m-\mu _{-}}^{\mu _{-}}\left(
z\right) \right\vert ^{2}dz,
\end{equation}%
which looks formally exactly the same as (\ref{exHO}) with the difference
that $\mu _{-}$ is given by the expression in (\ref{mumu}).

\section{A P\"{o}schl-Teller potential in disguise}

In the previous sections we observed that simple models on a noncommutative
space may lead to more unexpected solvable potential systems when expressed
in terms of the standard canonical variables and a subsequent
transformation. We may also reverse the question and explore which type of
model on a noncommutative space one obtains when we start from a well-known
solvable potential in the standard canonical variables. For instance, we
wish to construct the widely studied P\"{o}schl-Teller potential \cite%
{Poschl:1933zz}. Since the transformations are difficult to invert, we use
trial and error and find that this indeed achieved when starting with the
Hamiltonian 
\begin{equation}
H_{(i)}=\frac{\beta }{2m}P_{(i)}^{2}+\frac{\hbar \omega \alpha }{2\check{\tau%
}}P_{(i)}^{-2}+\frac{m\omega ^{2}}{2}X_{(i)}^{2}+\frac{\hbar \omega \alpha }{%
2}+\frac{\beta }{2m\check{\tau}}\quad \text{for }i=1,2,3,4;\alpha ,\beta \in 
\mathbb{R}.
\end{equation}%
We note that this Hamiltonian can not be viewed as a deformation of a model
on a standard commutative space as it is intrinsically noncommutative, in
the sense that it does not possess a trivial commutative limit $\tau
\rightarrow 0$. Proceeding as in the previous subsections we find for the
representation $\Pi _{(1)}$ that the Schr\"{o}dinger equation in momentum
space is once more of the general form of (\ref{1}), with%
\begin{equation}
f(p)=\frac{m\hbar ^{2}\omega ^{2}}{2}(1+\check{\tau}p^{2})^{2},\quad
g(p)=-m\hbar ^{2}\omega ^{2}\check{\tau}p(1+\check{\tau}p^{2}),\quad h(p)=%
\frac{(1+\check{\tau}p^{2})(\alpha m\hbar \omega +\beta p^{2})}{2m\check{\tau%
}p^{2}}.
\end{equation}%
From equation (\ref{psiq}) we obtain now%
\begin{equation}
\psi (p)=\phi (p),\quad q=\sqrt{\frac{2}{\tau \hbar \omega }}\arctan \left( 
\sqrt{\check{\tau}}p\right) ,\quad
\end{equation}%
and as anticipated we compute a P\"{o}schl-Teller potential with the help of
equation (\ref{V}) 
\begin{equation}
V_{PT}(q)=\frac{\hbar \omega \alpha }{2}\csc ^{2}\left( q\sqrt{\frac{\hbar
\omega \tau }{2}}\right) +\frac{\beta }{2m\check{\tau}}\sec ^{2}\left( q%
\sqrt{\frac{\hbar \omega \tau }{2}}\right) .  \label{VPT}
\end{equation}%
Assuming now that the special function $F(w)$ in (\ref{2nd}) is a Jacobi
polynomial $P_{n}^{(a,b)}(w)$, with $n\in \mathbb{N}_{0}$, $a,b\in \mathbb{R}
$, we identify from its defining differential equation, see e.g. \cite{Grad}%
, the coefficient functions in (\ref{2nd}) as%
\begin{equation}
Q(w)=\frac{b-a-(2+a+b)w}{1-w^{2}}\qquad \text{and\qquad }R(w)=\frac{%
n(n+1+a+b)}{1-w^{2}}.
\end{equation}%
Then equation (\ref{EV}) is evaluated to%
\begin{eqnarray}
E-V_{PT}(q) &=&\frac{n\left( w^{\prime }\right) ^{2}(a+b+n+1)}{1-w^{2}}+%
\frac{\left( w^{\prime }\right) ^{2}\left[ w^{2}(a+b+2)+2w(a-b)+a+b+2\right] 
}{2\left( 1-w^{2}\right) ^{2}}  \notag \\
&&-\frac{\left( w^{\prime }\right) ^{2}\left[ b-a-w(a+b+2)\right] ^{2}}{%
4\left( 1-w^{2}\right) ^{2}}-\frac{3\left( w^{\prime \prime }\right) ^{2}}{%
4\left( w^{\prime }\right) ^{2}}+\frac{w^{\prime \prime \prime }}{2w^{\prime
}},
\end{eqnarray}%
with as yet unknown function $w(q)$ and constant $E$. As in the previous
section we\ assume again that the first term on the right hand side gives
rise to a constant, i.e. $\left( w^{\prime }\right) ^{2}/(1-w^{2})=c\in 
\mathbb{R}^{+}$, but this time we choose the solution $w(q)=\cos (\sqrt{c}q)$%
, which solves (\ref{E}) with the identifications%
\begin{equation}
E_{n}=\frac{\hbar \omega \tau }{2}(1+2n+a+b)^{2},\quad c=2\tau \omega \hbar
,\quad a_{\pm }=\pm \frac{1}{2}\sqrt{1+\frac{4\alpha }{\tau },}\quad b_{\pm
}=\pm \frac{1}{2}\sqrt{1+\frac{4\beta }{\tau ^{2}}}.  \label{EPT}
\end{equation}%
Computing $v(q)$ by means of (\ref{v}) we assemble everything into the
solution of the Schr\"{o}dinger equation involving $H_{(1)}(x,p)$ 
\begin{equation}
\psi _{n}(p)=\frac{1}{\sqrt{N_{n}}}p^{1/2+a_{+}}(1+\check{\tau}%
p^{2})^{-(1+a_{+}+b_{+})/2}P_{n}^{(a_{+},b_{+})}\left( \frac{1-\check{\tau}%
p^{2}}{1+\check{\tau}p^{2}}\right) .  \label{ps}
\end{equation}%
We have selected here $a_{+}$ and $b_{+}$ in order to implement the
appropriate boundary conditions $\lim\nolimits_{p\rightarrow \pm \infty
}\psi _{n}(p)=0$ together with $\psi _{n}(0)=0$. We note that the energy
eigenvalues are real and bounded from below as long as $\alpha >-\tau /4$
and $\beta >-\tau ^{2}/4$. The occurrence of exceptional points is due to
the ${\mathcal{PT}}$-symmetry breaking of the wavefunction $\psi _{n}(p)$
when $a_{+}$, $b_{+}\notin \mathbb{R}$.

Following the same procedure as in the previous subsections we find for the
remaining representations 
\begin{equation}
\tilde{H}_{(1)}(q)=\tilde{H}_{(2)}(q)=\tilde{H}_{(3)}(q)=\tilde{H}_{(4)}(q),
\end{equation}%
where $q$ is related to $p$ differently in each case. Converting between the
different variables and computing the relevant pre-factors as in the
previous subsection we then find 
\begin{eqnarray}
\psi _{(2)}(p) &=&\rho _{(1)}^{-1/2}\psi _{(1)}(p), \\
\psi _{(3)}(p) &=&\frac{1}{\sqrt{N_{n}}}\frac{\left[ 1-\cos (2p\sqrt{\check{%
\tau}}))\right] ^{\frac{1+a_{+}}{2}}\left[ \cos (2p\sqrt{\check{\tau}})+1)%
\right] ^{\frac{1+b_{+}}{2}}}{\sqrt{\sin (2p\sqrt{\check{\tau}})}}%
P_{n}^{(a_{+},b_{+})}\left[ \cos \left( 2p\sqrt{\check{\tau}}\right) \right]
,~~~~ \\
\psi _{(4)}(p) &=&\frac{1}{\sqrt{N_{n}}}p^{a_{+}+1/2}(1+\check{\tau}p^{2})^{%
\frac{2b_{+}-1}{4}}P_{n}^{(a_{+},b_{+})}\left( 1+2\check{\tau}p^{2}\right) ,
\end{eqnarray}%
for the energy eigenvalue (\ref{EPT}) where $p>0$ for $\Pi _{(2)}$, $-\pi /2%
\sqrt{\check{\tau}}\leq p\leq \pi /2\sqrt{\check{\tau}}$ for $\Pi _{(3)}$
and $-i/\sqrt{\check{\tau}}\leq p\leq i/\sqrt{\check{\tau}}$ for $\Pi _{(4)}$%
. Using the the orthogonality relation for the Jacobi polynomial,\footnote{%
\begin{equation*}
\dint\nolimits_{-1}^{1}(1-x)^{a}(1+x)^{b}P_{n}^{(a,b)}\left( x\right)
P_{m}^{(a,b)}\left( x\right) dx=\delta _{n,m}N_{n}~\text{\ \ \ \ \ for }%
\func{Re}a,\func{Re}b>-1
\end{equation*}%
with $N_{n}=\frac{2^{a+b+1}\Gamma \left( a+n+1\right) \Gamma \left(
b+n+1\right) }{n!\Gamma \left( a+b+n+1\right) \Gamma \left( a+b+2n+1\right) }
$.} we compute the metrics from (\ref{rho}) to $\rho _{(1)}(p)=-2\sqrt{%
\check{\tau}}(1+\check{\tau}p^{2})^{-1}$, $\rho _{(2)}(p)=1$, $\rho
_{(3)}(p)=-2\sqrt{\check{\tau}}$ and\ $\rho _{(4)}(p)=2i\sqrt{\check{\tau}}%
(1+\check{\tau}p^{2})^{1/2}$.

We also note that for representation $\Pi _{(4^{\prime })}$ we obtain the
same potential (\ref{VPT}) with $\csc ^{2}\rightarrow \func{csch}^{2}$, $%
\sec ^{2}\rightarrow -\func{sech}^{2}$ plus an overall constant, which is
once again unphysical in the sense of leading to an unbounded spectrum from
below.

Finally we compute the expectation values for some arbitrary function $%
F\left( P_{(i)},X_{(i)}\right) $ 
\begin{equation}
\left\langle \psi _{(i)}\right\vert F\left( P_{(i)},X_{(i)}\right) \left.
\psi _{(i)}\right\rangle _{\rho _{(i)}}=\frac{1}{N}\dint\nolimits_{-1}^{1}F%
\left[ \frac{z}{\sqrt{\check{\tau}(1-z^{2})}},i\hbar \sqrt{\check{\tau}%
(1-z^{2})}\partial _{z}\right] \left\vert P_{n}^{(a_{+},b_{+})}\left(
z\right) \right\vert ^{2}dz,
\end{equation}%
which is again the same for all four representations.

\section{Conclusions}

We have shown how different representations for the operators $X$ and $P$
obeying a generalized version of Heisenberg's uncertainty relation are
related to each other by the transformations outlined in section. We have
demonstrated their equivalence within the setting of three
characteristically different types of solvable models, a Hermitian one, a
non-Hermitian one and an intrisically noncommutative one. In all cases we
showed that an appropriate metric can be found such that expectation values
result to be representation independent. We provided an explicit formula for
this metric, involving the quantities computed in the first two steps of the
general procedure. The computations were carried out in momentum space, but
naturally the method works equally well in standard $x$-space. In both cases
the order of the differential equation imposes a limitation on the type of
models which may be considered.

For representation $\Pi _{(4^{\prime })}$ proposed in \cite{Castro:2011in}
we found that it does not lead to the uncertainty relations (\ref{one}) and
moreover that for the models investigated it always gives rise to unphysical
spectra which are not bounded from below. This suggests that the general
procedure of Jordan twists requires a mild modification as outlined in the
manuscript.

Clearly it would be interesting to extent this analysis to different types
of full three dimensional algebras for noncommutative spaces and investigate
alternative representations, such as for instance for those already reported
in \cite{DFG}.

\bigskip \noindent \textbf{Acknowledgments:} SD is supported by a City
University Research Fellowship. AF would like to thank John Madore and
Francesco Toppan for useful discussions. BK would like to thank Andreas Fring 
and City University London for kind hospitality and The University of Jijel  
for financial support.

\newif\ifabfull\abfulltrue


\end{document}